\documentclass[twocolumn,prb, showpacs]{revtex4}
\usepackage{graphicx}
\usepackage{dcolumn}
\usepackage{bm}

\begin{document}


\title{Li non-stoichiometry and crystal growth of untwinned 1D quantum spin system Li$_{x}$Cu$_{2}$O$_{2}$}

\author{H. C. Hsu$^{1,2}$}
\author{H. L. Liu$^2$}
\author{F. C. Chou$^{1,3}$}
 \email{fcchou@ntu.edu.tw}
\affiliation{%
$^1$Center for Condensed Matter Sciences, National Taiwan University, Taipei 10617, Taiwan}%
\affiliation{%
$^2$Department of Physics, National Taiwan Normal University, Taipei 116, Taiwan}%
\affiliation{%
$^3$National Synchrotron Radiation Research Center, HsinChu 30076, Taiwan}%

\date{\today}

\begin{abstract}
Floating-zone growth of untwinned single crystal of Li$_x$Cu$_2$O$_2$ with
high Li content of x $  \sim$ 0.99$ \pm $0.03 is reported.  Li content of Li$_{x}$Cu$_{2}$O$_{2}$
has been determined accurately through combined iodometric titration and thermogravimetric methods, which also ruled out the speculation of chemical disorder between Li and Cu ions.
The morphology and physical properties of single crystals obtained from slowing-cooling (SL) and floating-zone (FZ) methods are compared.  The floating-zone growth under Ar/O$_2$=7:1 gas mixture at 0.64 MPa produces large area of untwinned crystal with highest Li content, which has the lowest helimagnetic ordering temperature $\sim$19K in the Li$_x$Cu$_2$O$_2$ system.

\end{abstract}


\maketitle


Li$_{x}$Cu$_{2}$O$_{2}$ has a Cu-O chain structure formed
with edge-sharing CuO$_{4}$ plaquettes in the $\textit{ab}$-plane
while these chains are connected through CuO$_{2}$ dumbbells alone
the $\textit{c}$-direction.\cite{Berger1991} This compound is uniquely
composed of nearly equal amount of Cu$^{1+}$ and Cu$^{2+}$
simply from the consideration of charge balance, where
O-Cu-O dumbbells are formed with Cu$^{1+}$. The crystal structure
was initially misidentified as a tetragonal symmetry due its severe twinning
until it is refined with an orthorhombic symmetry of $\textbf{b}$ $\sim$ 2$\textbf{a}$.\cite{Berger1991,Hibble1990}
Helimagnetic ordering has been identified by
neutron scattering along the $\textit{b}$-direction with an incommensurate
propagation vector (0.5, $\xi$ = 0.174, 0), where the spin spiral plane is
proposed to lie in the $\textit{ab}$-plane with pitching angle of
 2$\pi\xi$ $  \sim$ 62.6$ ^{\circ} $,\cite{Masuda2004} although the existence of transverse spiral spin component in the bc-plane is confirmed by Seki \emph{et al.} later.\cite{Seki2008}  Competing quantum and classical spin periodicity has been explored by neutron scattering, but the proposed important role of intrinsic chemical disorder in this 1D quantum spin system has not been examined fully yet.\cite{Masuda2004}

The importance of Li$_{x}$Cu$_{2}$O$_{2}$ compound has regenerated
great interest in the condensed matter physics community after it is
identified as the first cuprate system with multiferroic behavior by Park \textit{et al.} \cite{Park2007}
Spontaneous electric polarization emerges below the helimagnetic ordering temperature around $ \sim $ 22K and the direction of polarization
can be changed by the applied field.
However, due to the low atomic number of Li which prevents precise Li content determination,
theoretical models that is applied to interpret the origin of multiferroic behavior is
based mostly on the as-grown slow-cooled plus high temperature quenched single crystal
LiCu$_{2}$O$_{2}$.\cite{Bush2004}  The nominal stoichiometry
lacks precise Li and Cu contents analysis, which makes the interpretation of
these phenomena complicated, especially when the level of Cu$ ^{2+} $ impurity spin is the key parameter on inter-chain interaction.\cite{Masuda2004,Moskvin2008}
In particular, the slow cooling growth method that accompanies high temperature quenching in the air produces crystals with severe twinning and potential chemical disordering due to the similar ionic size between Li$^+$ and Cu$^+$.
The correlation between ferroelectricity and magnetic ordering,
and the competition between classical and quantum spins are clearly the most
important questions in current 1-d spin $ \frac{1}{2} $ system.
In order to clarify issues from materials point of view which is vital on constructing
an accurate theoretical model, we present details of crystal growth,
Li/Cu content analysis and helimagnetic ordering transition analysis based on magnetic
susceptibility measurement in this letter.

Due to the low atomic number limitation, Li content cannot
be determined accurately using either Inductive Coupled Plasma
(ICP) or Electron Probe Microanalysis (EPMA) techniques.  Li content, which is closely connected to the Cu$^+$/Cu$^{2+}$ ratio, has never been addressed carefully with confidence before.  Exact Li and Cu content has been extracted convincingly by combined iodometric titration
and thermogravimetric analysis in this study.  Contrary to the previous estimate of copper deficiency
and the speculated chemical disorder between Li and Cu,\cite{Bush2004} we find the Cu content is close to 2 and Li content is found to be lower than 1 always, which strongly suggests the doped Cu$^{2+}$ quantum spins that bridge the the CuO$_2$ ribbons between layers are originated from Li deficiency.
Herein we report detailed chemical analysis, Li level modification and characterization on nominal LiCu$_{2}$O$_{2}$ crystals
grown from slow-cooling and floating-zone methods.  Direct correlation between Li non-stoichiometry and the helimagnetic ordering transition
temperature is implied from our current study.

\begin{figure}
\begin{center}
\includegraphics[width=3.5in]{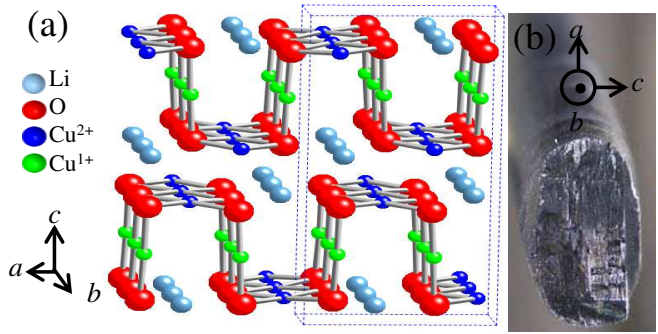}
\end{center}
\caption{\label{fig:fig1}(color online) (a)Crystal structure of LiCu$_{2}$O$_{2}$, where the magnetic Cu$ ^{2+} $ ion is shown in blue and the nonmagnetic Cu$^{1+}$ is in green.(b)A typical FZ crystal grown under Ar/O$_2$=7:1 gas mixture of 0.64 MPa. Note the crystal grows into an elliptical cross section with flat ab-plane after the first centimeter pulling and the growth direction is along the b-axis.}
\end{figure}


Floating-zone crystal (called FZ) has been grown from the feed rod
prepared with Li$_{2}$CO$_{3}$/CuO mixture of molar ratio 1.2 : 4, the cold-pressed feed rod of ~10 cm $\times$ $\oslash$6 mm is annealed at
850 $ ^{\circ} $C for 12 hours under the oxygen flow. The 20 percent excess of Li content is used to compensate for the high temperature Li vapor loss and the crystal growth is presumably through the congruent melt of the feed rod directly.
The optimum growth rate has been found to be 3 mm/hr and a 20 rpm rotation is maintained.  Various gas environment of different Ar/O2 ratios has been tested in order to achieve single phase growth with specific Li content.  Slow-cooled crystal (called SL) following recipe reported by Bush \textit{et al.} is prepared.\cite{Bush2004}
Final Li content can also been tuned through vacuum annealing starting from the as-grown FZ or SL crystals.  The crystals are sealed within an evacuated quartz tube and annealed at 400 and 600 $ ^{\circ} $C
for 6-24 hours respectively (called SL-(4)600 and FZ-(4)600).  There is white deposit generated inside the quartz tubing after the vacuum annealing, presumably due to the Li vapor loss which is strongly temperature dependent.  

Iodometric titration is performed to estimate the Cu$ ^{+} $/Cu$ ^{2+} $ ratio.  Since only the Cu$^{2+}$ has the oxidation power and the sample contains nearly half of the Cu$^+$ per formula unit, a two-step titration procedure is applied.  In the first step, single crystal sample
is ground and dissolved in 1M HCl, boiled for 5 min to ensure all Cu$ ^{+} $ is converted
into Cu$ ^{2+} $, and excess KI is added to the solution before titration
proceeds under N$_{2}$ flow.  While in the second step, titration is performed without the boiling procedure so that only the Cu$^{2+}$ portion is determined.  
The amount of Cu$^{2+}$ in the solution can be determined indirectly through the iodine content titrated with
0.1 M Na$ _{2} $S$ _{2} $O$  _{3}$ using deflecting point of the electric potential
as the indicator of titration ending point.  The complete reaction steps include
\begin{equation}
Cu ^{2+} + 3 I^{-} \rightarrow CuI + I _{2} + e^{-},
\label{eq:one}
\end{equation}
\begin{equation}
I_{2} + 2 S _{2} O _{3}  ^{2-} \rightarrow 2 I ^{-}  + S  _{4}O_{6} ^{2-}.
\label{eq:two}
\end{equation}

\noindent The Li content can be estimated from Li$_{1-z}$Cu$^{+}_{1-z}$Cu$^{2+}_{1+z}$O$_2$ formula after the total copper content is normalized to 2 per formula unit.

Thermogravimetric decomposition method is applied to determine the Li content and compare with that from titration.  Li$ _{1-z} $Cu$ _{2} $O$ _{2} $ can be decomposed into Li$ _{2} $CuO$ _{2} $ and CuO under oxygen flow as
\begin{equation}
Li _{1-z} Cu _{2} O _{2} \rightarrow \frac{1-z}{2} Li_{2}CuO_{2} + \frac{3+z}{2}CuO.
\label{eq:three}
\end{equation}

\noindent Li content can be calculated from the weight gain after the sample is
converted into Li$ _{2} $CuO$ _{2} $ and CuO completely.  Since titration method requires much more sample in powder form to increase the accuracy, TGA decomposition method has been used to determine the Li content of single crystal effectively in this report, although it has a larger error bar $\pm$0.03 and is systematically higher than that from titration by about 0.02.

Magnetic susceptibility is measured with SQUID magnetometer (Quantum Design MPMS-XL) with a field of 1 kOe for magnetic field applied along and perpendicular to the ab-plane.  Since the spin susceptibility of helimagnetic ordering varies smoothly
comparing with the conventional FM or AFM ordering, d$ \chi $/dT
instead of $ \chi $ versus temperature is used to identify the magnetic phase transition.

\begin{table}
\caption{\label{tab:table-FZ}Growth conditions and products for floating-zone growth of Li$_{x}$Cu$_{2}$O$_{2}$}
\begin{tabular}{cccc}
 \hline
Sample & Li content (TGA) & Ar:O$_{2}$ & compound(s) \\
 \hline
SL & 0.87 $\pm$ 0.03 & air & Li$_{x}$Cu$_{2}$O$_{2}$\\
FZ-1 &  & 5:1 & Li$_{x}$Cu$_{2}$O$_{2}$ $+$ $\sim$ 2 \%LiCu$_{3}$O$_{3}$\\
FZ-2 & 0.99 $\pm$ 0.03 & 7:1 & Li$_{x}$Cu$_{2}$O$_{2}$ \\
FZ-3 & 0.95 $\pm$ 0.03 & 10:1 & Li$_{x}$Cu$_{2}$O$_{2}$ \\
FZ-4 & 0.84 $\pm$ 0.03 & Ar & Li$_{x}$Cu$_{2}$O$_{2}$ \\
 \hline
\end{tabular}
\end{table}

\begin{table}
\caption{\label{tab:table-Titration}Li content of Li$ _{1-z} $Cu$ _{1-z} ^{+}$Cu$ _{1+z} ^{2+}$O$ _{2} $ derived from titrated Cu$ ^{2+} $/Cu$ ^{+} $ ratio with total Cu normalized to 2.}
\begin{tabular}{ccccc}
 \hline
Titration No. & 1-z(FZ) & Cu$^{2+}$/Cu$^{+}$(FZ) & 1-z(SL) &  Cu$^{2+}$/Cu$^{+}$(SL) \\
 \hline
1 & 0.96 & 52 /48 \% & 0.83 &  58.5 /41.5 \% \\
2 & 0.96 & 52 /48 \% & 0.83 &  58.5 /41.5 \% \\
3 & 0.96 & 52 /48 \% & 0.84 &  58 /42 \% \\
4 & 0.96 & 52 /48 \% & 0.84 &  58 /42 \% \\
5 & 0.95 & 52.5 /47.5 \% & 0.84 &  58 /42 \% \\
 \hline
\end{tabular}
\end{table}
					

We have tested the floating-zone growth under various growth conditions in order to control the Li content and summarized in Table~\ref{tab:table-FZ}. 
The higher gas pressure of 0.64 MPa reduces Li vapor loss significantly, although the Li content and impurity level are also closely related to the oxygen partial pressure. 
Single phase crystal of highest Li content is obtained under Ar/O$_2$=7:1 gas environment.  

\begin{figure}
\begin{center}
\includegraphics[width=3.5in]{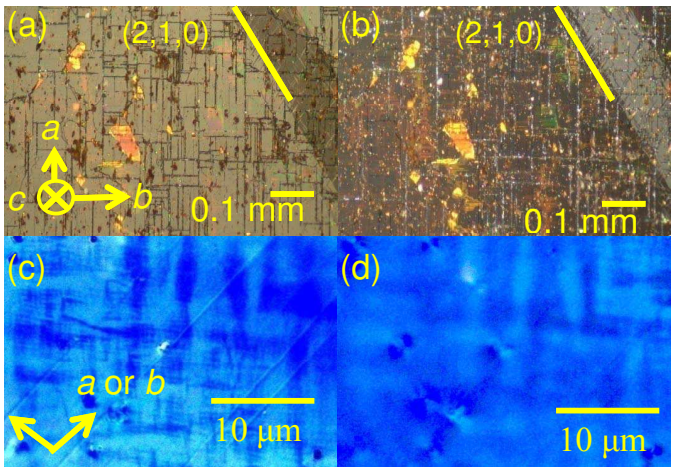}
\end{center}
\caption{\label{fig:fig2}(color online) (a) and (b) are optical microscope images of FZ crystal taken with de-polarized light of 90 $^\circ$ difference. Similarly (c)-(d) are taken from SL crystal.}
\end{figure}

Figure~\ref{fig:fig1}(b) shows the typical FZ crystal grown under Ar/O$_2$=7:1 gas mixture of 0.64 MPa.  The FZ crystal shows flat surface after the first $\sim$ 1 cm pulling and the flat surface is verified by X-ray diffraction to be the ab-plane, i.e. the plane that contains CuO$_4$ plaquettes as shown in Fig.~\ref{fig:fig1}(a).  Crystal surface images taken by polarized light are displayed in Fig.~\ref{fig:fig2} for both FZ and SL crystals.  Note the area of the dark-light contrast of Fig.~\ref{fig:fig2}(a)-(b) for FZ crystal and Fig.~\ref{fig:fig2}(c)-(d) for SL crystal have different scales, where FZ crystal shows large untwinned area $\sim$ 3 $\times$ 3 mm$^2$, which is more than 100 times larger than the severely twinned SL crystal.  There is only stripe like second domain observed sporadically on the FZ crystal surface as shown in Fig.~\ref{fig:fig1}(b).  The FZ crystal growth direction follows closely to the edge-shared CuO$_4$ plaquettes, i.e. the spiral spin chain b-axis direction.\cite{Masuda2004}  Mirror plane of (210) which is parallel to the domain boundary is clearly indicated by the highlighted lines in Fig.~\ref{fig:fig2}(a)-(b), in agreement with the proposed \textbf{a} $\sim$ 2\textbf{b} structure model.  

The Cu$^{+}$/Cu$^{2+}$ ratio can be determined by combined iodometric titration and TGA decomposition techniques accurately.  However, the Cu and oxygen contents should be addressed first while we must use an accurate formula unit in the calculation, in particular before the possibility of Li$^+$/Cu$^+$ chemical disorder is not ruled out yet.\cite{Bush2004}  We tentatively assign the formula unit of the sample to be LiCu$_2$O$_2$, calculate the predicted total copper content and compare with that obtained from titration.  The titrated total copper content is obtained after a boiling procedure, as described in the Experimental section, in order to convert all copper ions into Cu$^{2+}$ state.
The copper content based on LiCu$_{2}$O$_{2}$ formula
weight and its predicted titration level is compared with the experimental results as shown in Table~\ref{tab:table-Cu}. We find the Cu content
which is converted from the titrated Na$ _{2} $S$ _{2} $O$  _{3}$ volume is nearly the same
as the calculated values from predicted level using LiCu$_{2}$O$_{2}$ formula,
which implies the assumed copper content of 2 is correct.  In addition, the assumption of Cu to be 2 per formula unit holds true for crystals grown from both FZ and SL methods, which is against the possibility of Li$^+$/Cu$^+$ chemical disorder while two different degrees of quenching have been applied.
The consistency displayed in Table ~\ref{tab:table-Cu} also supports the assumption
about oxygen content to be close to 2 within 2 \% error.
ICP analysis on NaCu$ _{2} $O$ _{2} $ shows that Cu
content to be 2$ \pm $0.02 also while it has a similar structure
with a slightly lower magnetic transition temperature only.\cite{Capogna2005}
In addition, oxygen non-stoichiometry can be reasonably
ruled out due to its significant stability between 890-1050 $ ^{\circ} $C
as reported by Bush \textit{et al.} \cite{Bush2004}

\begin{table}
\caption{\label{tab:table-Cu}Comparison of Cu content obtained from iodometric titration and predicted calculation}
\begin{tabular}{ccc}
 \hline
Sample& Cu by titration (mol)& Cu from calculation (mol)\\
 \hline
SL (15 mg) & (1.81 $\pm $ 0.03) $ \times $ 10$ ^{-4} $ & (1.8068 $\pm $ 0.014) $ \times $ 10$ ^{-4} $\\
FZ (10 mg) & (1.235 $\pm$ 0.02) $ \times $ 10$ ^{-4} $ & (1.2045 $\pm $ 0.012) $ \times $ 10$ ^{-4} $\\
 \hline
\end{tabular}
\end{table}

\begin{figure}
\begin{center}
\includegraphics[width=3.5in]{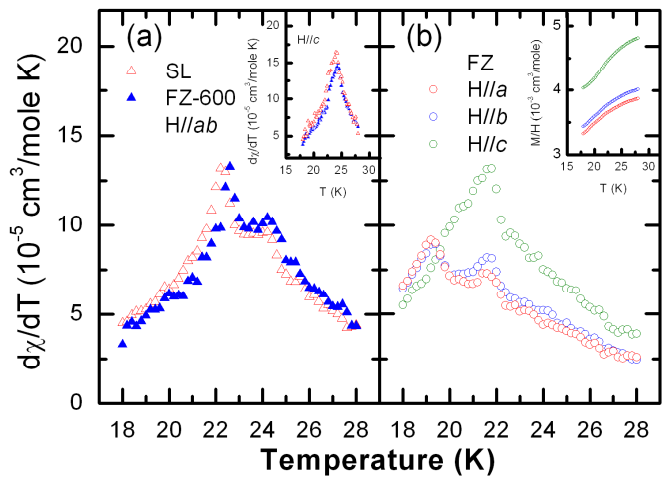}
\end{center}
\caption{\label{fig:fig3}(color online) (a) Temperature dependent d$\chi$/dT of Li$_{x}$Cu$_{2}$O$_{2}$ for samples prepared from slow cooling (SL) and 600 $^\circ$C vacuum annealed FZ sample (FZ-600). H$\parallel\emph{c}$ data are shown in the inset. (b) d$\chi$/dT for the as-grown crystal FZ with highest x $\sim$0.99. The original data $\chi$(T) are shown in the inset.}
\end{figure}

Titration results that determine the Cu$ ^{+} $/Cu$ ^{2+} $ ratio for two samples prepared by FZ and SL methods are summarized in Table~\ref{tab:table-Titration}.
We can easily convert the titrated Cu$ ^{+} $/Cu$ ^{2+}$ ratio to Li content
through charge neutrality requirement with the predicted  Li$ _{1-z} $Cu$ _{1-z} ^{+}$Cu$ _{1+z} ^{2+}$O$ _{2} $ formula, where total copper content is normalized to 2.  Li content is obtained to be 0.838 $  \pm$ 0.01 and 0.964 $ \pm $ 0.015 for SL and FZ samples respectively.  We find the Li content of SL growth is highly reproducible, the Li content near 0.83 could have a special stability, in particular its magnetic transition represented by the d$\chi$/dT peak shape (see Fig.~\ref{fig:fig3}(a) below) is sharp and reproducibly similar to those reported in the literature.\cite{Seki2008}
Li$_x$Cu$_2$O$_2$ is able to decompose into Li$_2$CuO$_2$ and CuO under oxygen environment and the Li content can also be calculated following Eq.~(\ref{eq:three}) described above. 
Li content determined by TGA shows larger error bar and is
consistently higher than that obtained from titration by $ \sim $ 0.02.
Although titration provides values of higher consistency and accuracy,
significantly larger amount of sample in polycrystalline form is required,
the Li content values reported in this paper are based mainly on TGA results because the limited sample size of annealed single crystal samples.  

Since the spin susceptibility of helimagnetic ordering varies smoothly
comparing with that of the conventional FM or AFM ordering, d$ \chi $/dT
instead of $ \chi $ versus temperature for FZ samples are
summarized in Fig.~\ref{fig:fig3}. We note the sample of FZ-600, which is prepared from the as-grown FZ crystal of x $\sim$0.99 after 600$^\circ$C vacuum annealing, has identical Li content of $\sim$0.84 and magnetic profile to that of the as-grown SL sample shown in Fig.~\ref{fig:fig3}(a), which suggests physical properties must be critically determined by the Li content.  Similar to what Seki \emph{et al.} have found,\cite{Seki2008} there are two anomalies near $\sim$ 22.5K and 24K for both $\emph{ab}$- and $\emph{c}$-directions, although 22.5K is dominant for H$\parallel$$\emph{ab}$ and 24K is dominant for H$\parallel$$\emph{c}$.  Very importantly, spontaneous electric polarization is found only below 22.5K.\cite{Seki2008}  Since we can measure the spin susceptibility along a- and b-axes independently because of the sizable untwinned crystal, susceptibility data along three crystal axes for the as-grown FZ crystal (x $\sim$0.99) are shown in Fig.~\ref{fig:fig3}(b).  Similar two-peak profile near 19.2 and 21.7 K are found, which are about 2 degrees lower than those found in the SL sample and in the literature of identical
growth method.\cite{Masuda2004,Seki2008,Park2007}  The FZ sample of the highest Li content shows the lowest transition temperature, which is the lowest magnetic transition temperature found in this system so far.  We find there is no significant difference between $\emph{a}$- and $\emph{b}$-directions and the transition near 19.2K is less pronounced as indicated by the lower d$\chi$/dT value.  According to the similar d$\chi$/dT profiles between $\emph{a}$- and $\emph{b}$-axes for the FZ crystal of the highest Li content, it would be reasonable to consider spin spirals are in the $\emph{ab}$-plane in a classical picture, but further evidence is required to make a definite conclusion.
 It would be important to compare whether helimagnetic spiral modulation vectors are also different for the FZ and SL crystals.  In addition, the existence of spontaneous electric polarization in the Li fully filled crystal must also be examined.

The Li deficiency induced Cu$^{2+}$ impurity spins could be strongly correlated to the  helimagnetic ordered spins within CuO$_2$ chains.  Such strong impact of impurity spins to the host ordered low dimensional system has been observed in quasi-1d AF CaCu$_2$O$_3$ also.\cite{Goiran2006}	In particular, similar incommensurate ordering of Dy moments with the same
periodicity as the Mn spiral ordering has been observed in the multiferroic DyMnO$_3$.\cite{Prokhnenko2007}  We speculate the highly reproducible phase of particular stability from slow cooling growth has a stoichiometry close to Li$_{0.83}$Cu$_{2}$O$_{2}$ always, and the $\sim$17 \% Li$ ^{+} $ deficiency level implies
the introduced Cu$ ^{2+} $ impurity could be directly
correlated to the incommensurate ($\xi$ = 0.174) spiral modulation along
the chain \textbf{b}-direction.\cite{Masuda2004,Seki2008}
While each missing Li$ ^{+} $ ion would convert one Cu$ ^{+}$ to Cu$ ^{2+} $
from the O-Cu-O dumbbell that bridges adjacent CuO$ _{2} $ edge-sharing chains,
the impurity Cu$ ^{2+} $ would support the nonrelativistic picture of
multiferroicity model naturally as proposed by Moskvin \textit{et al.},
without even the need of founding their theory on the assumption of Cu deficiency
and Li excess.\cite{Moskvin2008}  On the other hand, the assumption of direct correlation among Li deficiency, Cu$^{2+}$ impurity, helimagnetic ordering, and spontaneous electric polarization would require further rigorous cross checking.  A complete study on spiral modulation vector and electric polarization based on the whole series of Li$_x$Cu$_2$O$_2$ is underway.


In conclusion, we have presented floating-zone growth of large untwinned Li$_x$Cu$_2$O$_2$ single crystal with controlled Li deficiency level from 0.75 $\leq$ x $\leq$ 0.99.  While Li and Cu content is crucial in this compound but most
chemical analysis failed to decide Li content, current combined titration
and TGA studies provide a reliable way to determine the Li and Cu content with confidence. Our study suggests the actual chemical composition should be Li$ _{1-z} $Cu$ _{1-z} ^{+}$Cu$ _{1+z} ^{2+}$O$ _{2} $ and Cu$^{2+}$ impurity spins are introduced by Li$^+$ deficiency along the O-Cu-O dumbbells $\parallel$ $\emph{c}$-direction.  These results should pave the way to a better understanding on this intriguing 1D quantum spin system both theoretically and experimentally.

FCC acknowledges the support from National Science Council of Taiwan under project number NSC-95-2112-M-002.
\appendix


\begin{thebibliography}{99}

\bibitem{Berger1991}
R. Berger, A. Meetsma, S. van Smaalen, and M. Sundberg, J. Less-Common Met. \textbf{175}, 119 (1991).

\bibitem{Hibble1990}
S. J. Hibble, J. Kohler, and A. Simon, J. Solid State Chem. \textbf{88}, 534 (1990).

\bibitem{Masuda2004}
T. Masuda, A. Zheludev, A. Bush, M. Markina, and A. Vasiliev, Phys. Rev. Lett. \textbf{92}, 177201 (2004).

\bibitem{Seki2008}
S. Seki, Y. Yamasaki, M. Soda, M. Matsuura, K. Hirota, and Y. Tokura, Phys. Rev. Lett. \textbf{100}, 127201 (2008).

\bibitem{Park2007}
S. Park, Y. J. Choi, C. L. Zhang, and S.-W. Cheong, Phys. Rev. Lett. \textbf{98}, 057601 (2007).

\bibitem{Bush2004}
A. A. Bush, K. E. Kamentsev, and E. A. Tishchenko, Inorg. Mater. \textbf{40}, 44 (2004).

\bibitem{Moskvin2008}
A. S. Moskvin, Y. D. Panov, and S.-L. Drechsler, arXiv:0801.1975v1 (2008).

\bibitem{Capogna2005}
L. Capogna, M. Mayr, P. Horsch, M. Raichle, R. K. Kremer, M. Sofin, A. Maljuk, M. Jansen, and B. Keimer, Phys. Rev. B \textbf{71}, 140402(R) (2005).

\bibitem{Goiran2006}
M. Goiran, M. Costes, J. M. Broto, F. C. Chou, R. Klingeler, E. Arushanov, S.-L. Drechsler, B. Buchner, and V. Kataev, New J. Phys. \textbf{8}, 74 (2006).

\bibitem{Prokhnenko2007}
O. Prokhnenko, R. Feyerherm, E. Dudzik, S. Landsgesell, N. Aliouane, L. C. Chapon, and D. N. Argyriou, Phys. Rev. Lett. \textbf{98}, 057206 (2007)


\end{thebibliography}
\end{document}